\documentclass[runningheads]{llncs}

\usepackage[ruled,vlined]{algorithm2e}
\usepackage{array}
\usepackage{amsfonts}
\usepackage{amsmath}
\usepackage{amssymb}
\usepackage{color}
\usepackage{enumitem}
\usepackage[mathscr]{eucal}
\usepackage{graphicx,epstopdf}
\usepackage{hyperref}
\usepackage{latexsym}
\usepackage[numbers]{natbib}
\usepackage{newtxmath}
\usepackage{newtxtext}
\usepackage{srcltx} 
\usepackage{subcaption}
\usepackage[table]{xcolor} 

\graphicspath{ {figures/} }

\newcommand{\norm}[1]{\left\lVert#1\right\rVert}

\newcommand{\ie}{\emph{i.e.,}\xspace}

\newcommand{\etal}{\emph{et~al.}\xspace}
\newcolumntype{x}[1]{>{\raggedright\arraybackslash}p{#1}}

\SetCommentSty{mycommfont}

\makeatletter 
\g@addto@macro{\@algocf@init}{\SetKwInOut{Output}{Output}} 
\makeatother

\begin{document}

\title{Image Analysis Enhanced Event Detection from Geo-tagged Tweet Streams}

\author{Yi Han\orcidID{0000-0001-6530-4564} \and
Shanika Karunasekera\orcidID{0000-0001-7080-5064} \and
Christopher Leckie\orcidID{0000-0002-4388-0517}
}

\authorrunning{Y. Han et al.}

\institute{School of Computing and Information Systems, The University of Melbourne\\
\email{\{yi.han, karus, caleckie\}@unimelb.edu.au \\
}}

\maketitle

\begin{abstract}
Events detected from social media streams often include early signs of accidents, crimes or disasters. Therefore, they can be used by related parties for timely and efficient response. Although significant progress has been made on event detection from tweet streams, most existing methods have not considered the posted images in tweets, which provide richer information than the text, and potentially can be a reliable indicator of whether an event occurs or not. In this paper, we design an event detection algorithm that combines textual, statistical and image information, following an unsupervised machine learning approach. Specifically, the algorithm starts with semantic and statistical analyses to obtain a list of tweet clusters, each of which corresponds to an event candidate, and then performs image analysis to separate events from non-events---a convolutional autoencoder is trained for each cluster as an anomaly detector, where a part of the images are used as the training data and the remaining images are used as the test instances. Our experiments on multiple datasets verify that when an event occurs, the mean reconstruction errors of the training and test images are much closer, compared with the case where the candidate is a non-event cluster. Based on this finding, the algorithm rejects a candidate if the difference is larger than a threshold. Experimental results over millions of tweets demonstrate that this image analysis enhanced approach can significantly increase the precision with minimum impact on the recall.
\end{abstract}

\section{Introduction}\label{sec:intro}
While social media, especially Twitter, has gained growing popularity over the past decade, it has also become a new source of news---events detected from social media streams often contain early signs of accidents, crimes or disasters. Therefore, they can provide valuable information for related parties to take timely and efficient responses.

Although event detection from tweet streams has been extensively studied, most existing methods still suffer from relatively high false positive and false negative rates, especially for unsupervised machine learning approaches. These algorithms normally rely on semantic, spatial, temporal and frequency information. Images, on the other hand, have rarely been considered yet. Compared with text, especially short posts like tweets, images often provide richer information and potentially can help discover the occurrence of an event.

In this paper, we design an unsupervised event detection algorithm that utilises images in addition to textual and statistical information. The core idea is that when an event occurs, the images posted in the surrounding area are likely to be similar/correlated. Therefore, if we use part of them to train an autoencoder, and keep the rest as the test instances, the reconstruction errors of the training and test images should be close. However, when no event happens, the images posted in a certain region are likely to be more diverse, and hence the reconstruction errors of the test instances will be much higher than those of the training instances, as the autoencoder has not seen similar images before. Based on this idea, the algorithm uses the ratio between the mean reconstruction errors of the test and training images as an additional criterion to further decrease the false positive rate for event detection. Note that since image analysis is relatively expensive, it is only performed at the last step, after the semantic and statistical analyses are finished, which follow a similar approach to~\cite{han_multi-spatial_2019} with several improvements. In addition, considering that the posted images are normally limited, the algorithm randomly generates the same number of crops for each of them, and trains the autoencoder on the snippets. 

In summary, the main contributions of this paper include:

\begin{itemize}
    \item We analyse images posted in both event and non-event tweet clusters based on the reconstruction errors of autoencoders, and demonstrate that when an event occurs, the images are more coherent (Section~\ref{subsec:quantitative});
    \item We utilise this finding and propose an image analysis enhanced event detection algorithm from tweet streams. It should be emphasised that although we integrate image analysis with a specific existing method~\cite{han_multi-spatial_2019}, the analysis is generic and can be incorporated with other event detection schemes as well (Section~\ref{subsec:algo});
    \item We conduct experiments on multiple tweet datasets, and demonstrate that this unsupervised, image analysis enhanced approach can significantly increase the precision without any impact on the recall (Section~\ref{sec:experiment}).
\end{itemize}

The remainder of this paper is organised as follows: Section~\ref{sec:detection} specifies the event detection problem, and introduces the image analysis enhanced algorithm; Section~\ref{sec:experiment} presents the experimental verification; Section~\ref{sec:related} overviews previous work on event detection; and Section~\ref{sec:conclusions} concludes the paper and gives directions for future work.

\section{Image Analysis Enhanced Event Detection}\label{sec:detection}
In this section, we start with a brief definition of the event detection problem from geo-tagged tweet streams, then introduce in detail how image analysis is performed, and how it is integrated with semantic and statistical analyses.

\subsection{Autoencoder based Image Analysis}
We study the event detection problem defined as follows: given a tweet stream \(T = \{t_{1}, t_{2}, ..., t_{n}\}\) from a certain region, and a query window \(W = \{t_{n-m+1}, t_{n-m+2}, ..., t_{n}\}\) (\(m\) is the number of tweets in \(W\)) that represents currently observed tweets, the aim is to identify a set of tweets \(T_{i} \subseteq W\) that are associated with an event, e.g., an accident, a disaster or protest, as close to where and when the event occurs as possible.

A common type of solution to the above problem takes the clustering based approach~\cite{abdelhaq_eventweet:_2013,hasan_real-time_2019,li_tedas:_2012,xie_topicsketch:_2016,zhang_geoburst+:_2018,zhang_triovecevent:_2017,zhang_geoburst:_2016}, which generates a list of event candidates by clustering the tweets according to their semantic, spatial and temporal information, and then removes non-event clusters via supervised or unsupervised methods. In this work, we focus on how image analysis can be used to enhance the second step.

Specifically, suppose that a set of images, \(\mathit{IM} =\{im_{1}, im_{2}, ..., im_{k}\}\), are extracted from an event candidate, \ie a cluster of tweets that are semantically coherent, and geographically and temporally close, \(\mathit{IM}\) is divided into two subsets \(\mathit{IM}_{train} \subset \mathit{IM}\), \(\mathit{IM}_{test} = \mathit{IM} \setminus \mathit{IM}_{train}\), which are the training and test datasets, respectively. For each image \(im_{i} \in \mathit{IM}\), \(c\) random crops of the same size are generated, \(\{im_{ij},\ j=1, 2, ..., c\}\), and \(\{im_{ij}\ |\ im_{i} \in \mathit{IM}_{train}\}\) are used to train a convolutional autoencoder, while \(\{im_{ij}\ |\ im_{i} \in \mathit{IM}_{test}\}\) are kept as the test instances.

As mentioned in the introduction, when an event occurs the images in \(\mathit{IM}\) are likely to be similar, and hence the reconstruction errors of \(\{im_{ij}\ |\ im_{i} \in \mathit{IM}_{train}\}\) should be close to those of \(\{im_{ij}\ |\ im_{i} \in \mathit{IM}_{test}\}\). On the other hand, when there is not any event the difference in the reconstruction errors between the training and test instances should be much larger. Therefore, we propose to quantify the coherence of the images in a cluster, and use that as a metric to detect and remove non-event clusters.

\subsection{Quantitative Study}\label{subsec:quantitative}
In order to validate the above idea, we collected (part of) the posted images in the following three Twitter datasets:

\begin{itemize}
    \item Dataset shared by the authors of~\cite{zhang_geoburst:_2016}, which includes 9.5 million geo-tagged tweets from New York between 1 August, 2014 and 30 November 2014---617K images are retrieved from it;
    \item All geo-tagged tweets from Los Angeles between 9 February and 22 February 2019, with a size of 13.2K---20K images are retrieved from it;
    \item All geo-tagged tweets from Sydney between 12 February and 5 April 2019, with a size of 28.4K---16K images are retrieved from it.
\end{itemize}

For each dataset, we first perform semantic and statistical analyses using the method in~\cite{han_multi-spatial_2019} (more details are given in the next subsection) to obtain a list of event candidates. If a candidate contains at least three images, we then (1) randomly generate 500 crops of size \(32 \times 32\) for each image---there are normally a limited number of images within each cluster, and they are insufficient for the training of an autoencoder; (2) use two-thirds of the crops to train a convolutional autoencoder, and keep the rest as the test data. Note that all the 500 crops of an image are either in the training or test dataset. In addition, we also notice that if a considerable part of an image is about human beings, the image is often quite different from the rest even if there is an event. For example, during a sports game or a concert, while the focus of most images is the court or the stadium, selfie images are likely to be very different and hence cause false negatives. Therefore, images of this type are excluded in the analysis (see Section~\ref{subsec:setup} for more details), \ie each cluster needs to have at least three non-human images in order to be analysed; (3) compare the mean, median and variance of the reconstruction errors (REs, \(RE(x) = \norm{x - x^\prime}^2\), where \(x\) and \(x^\prime\) are the input and output of the autoencoder, respectively) for the training and test instances, and calculate the ratios of \(\frac{mean(RE_{test})}{mean(RE_{train})}\), \(\frac{median(RE_{test})}{median(RE_{train})}\), and \(\frac{var(RE_{test})}{var(RE_{train})}\), where \(RE_{train}\) and \(RE_{test}\) represent the set of training and test REs, respectively.

\begin{figure}[t!]
\centering
\begin{subfigure}{0.4\columnwidth}
  \centering
  \includegraphics[width=\columnwidth]{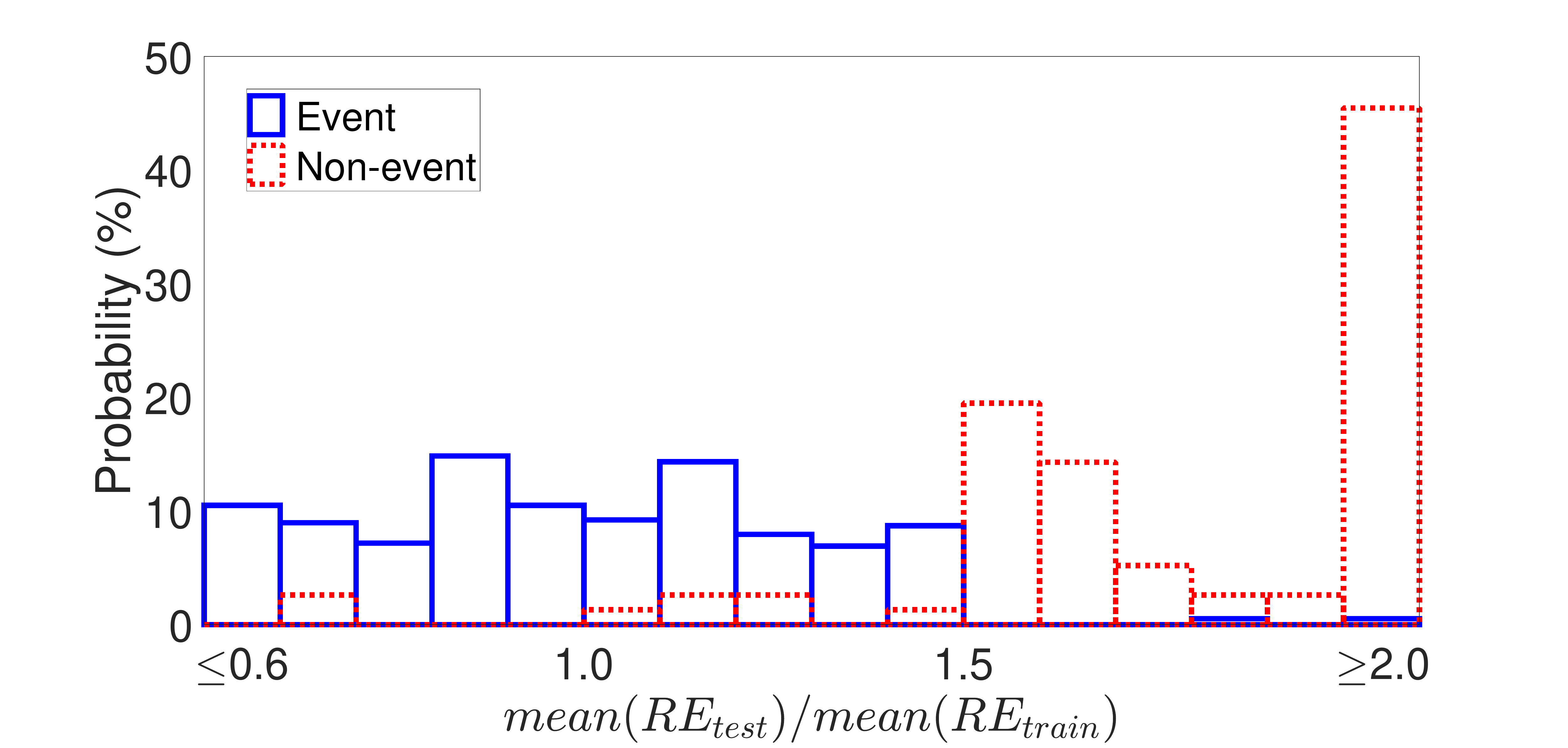}
  \caption{Mean REs: New York.}
  \label{figure_r_ny}
\end{subfigure}%
\begin{subfigure}{0.4\columnwidth}
  \centering
  \includegraphics[width=\columnwidth]{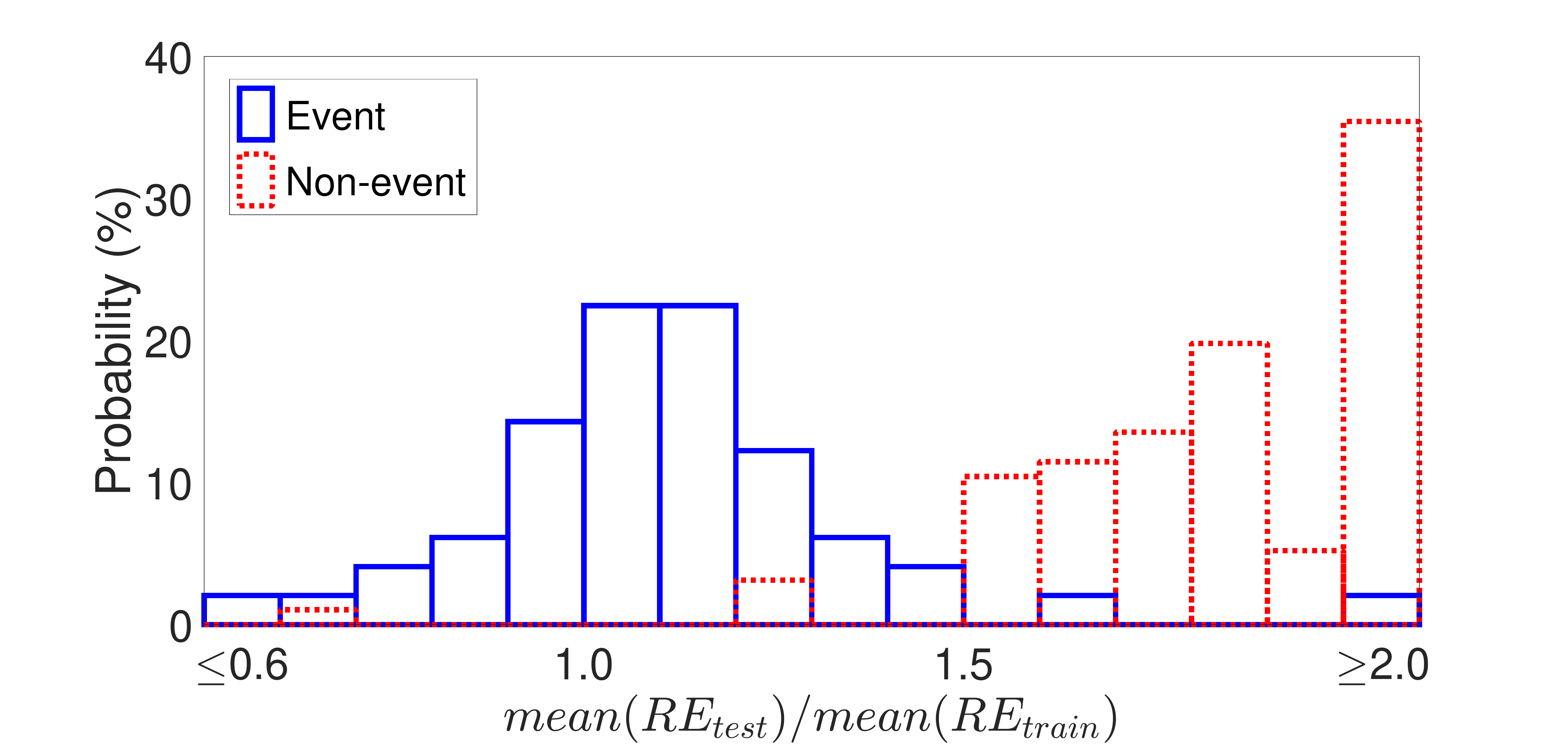}
  \caption{Mean REs: LA \& Sydney.}
  \label{figure_r_la_syd}
\end{subfigure}
\begin{subfigure}{0.4\columnwidth}
  \centering
  \includegraphics[width=\columnwidth]{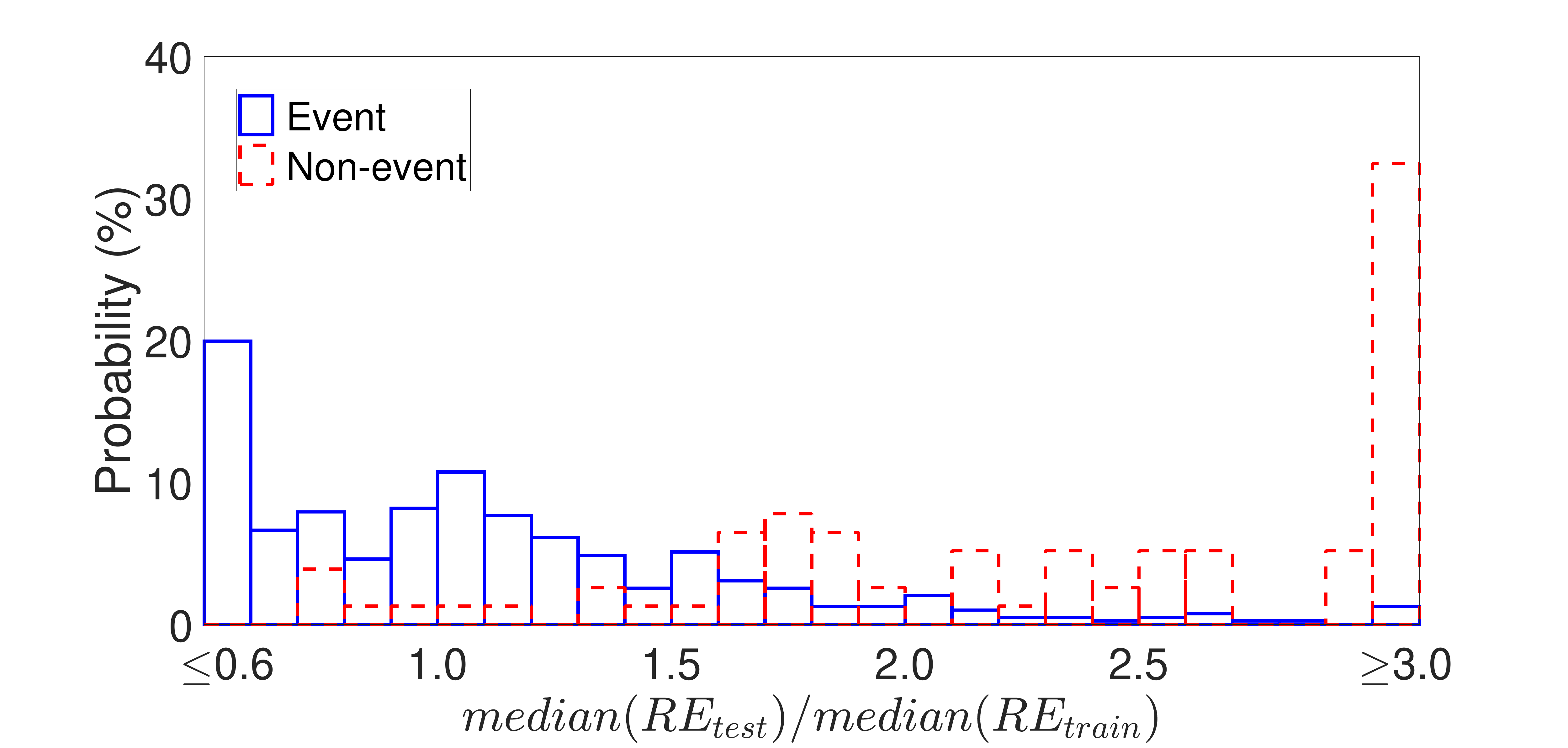}
  \caption{Median REs: New York.}
  \label{figure_median_ny}
\end{subfigure}%
\begin{subfigure}{0.4\columnwidth}
  \centering
  \includegraphics[width=\columnwidth]{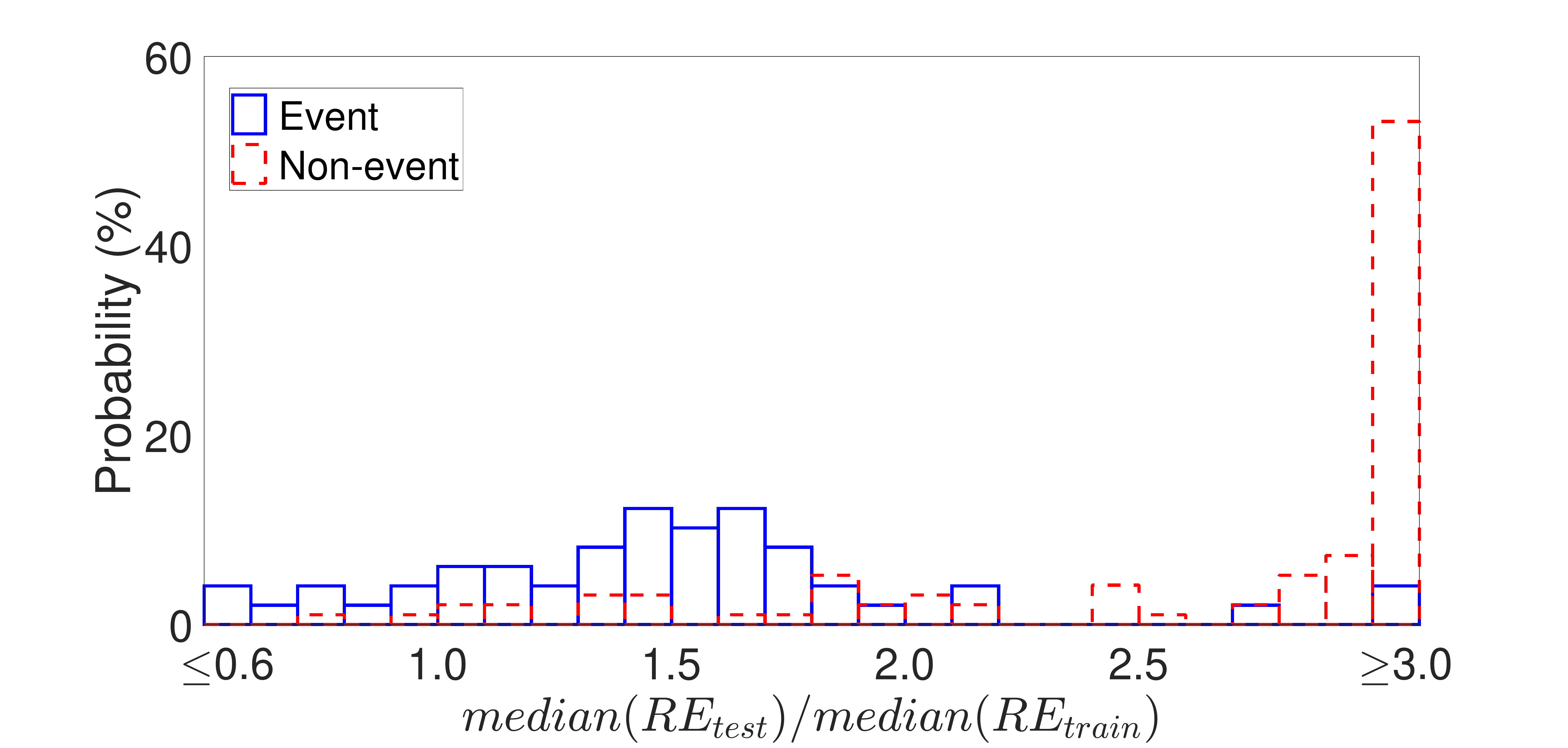}
  \caption{Median REs: LA \& Sydney.}
  \label{figure_median_la_syd}
\end{subfigure}
\begin{subfigure}{0.4\columnwidth}
  \centering
  \includegraphics[width=\columnwidth]{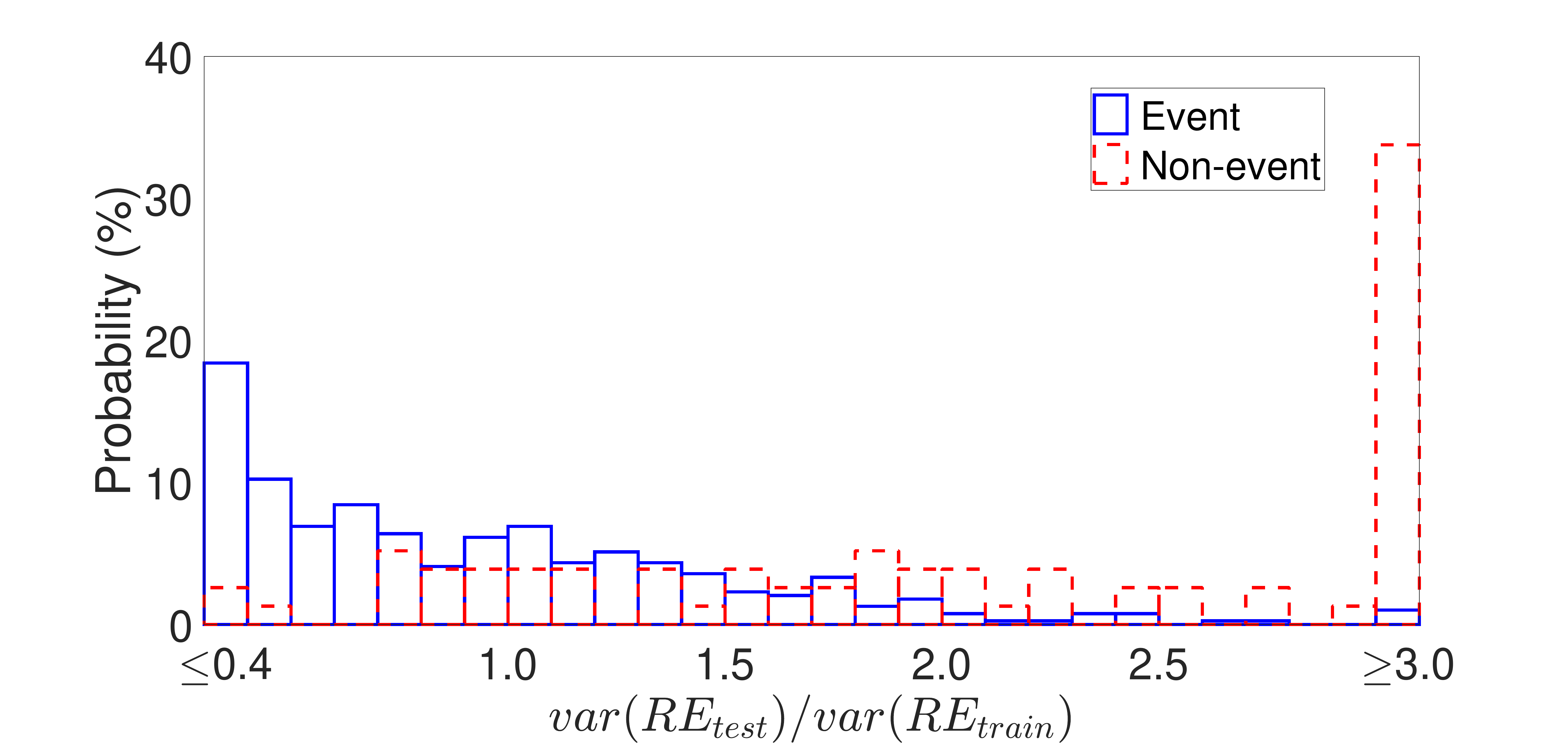}
  \caption{Variance of REs: New York.}
  \label{figure_var_ny}
\end{subfigure}%
\begin{subfigure}{0.4\columnwidth}
  \centering
  \includegraphics[width=\columnwidth]{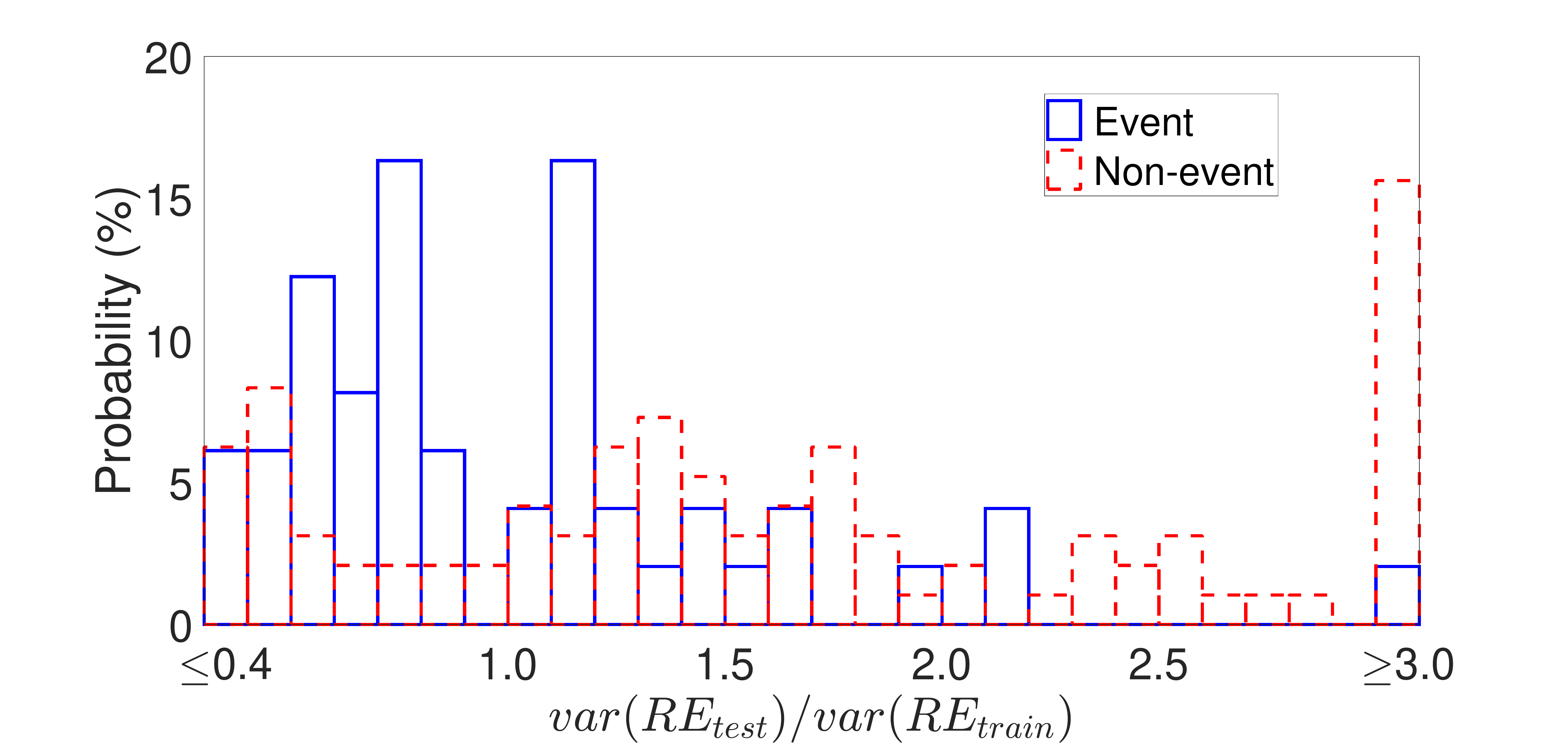}
  \caption{Variance of REs: LA \& Sydney.}
  \label{figure_var_la_syd}
\end{subfigure}
\caption{Probability distributions of \(\frac{mean(RE_{test})}{mean(RE_{train})}\), \(\frac{median(RE_{test})}{median(RE_{train})}\), and \(\frac{var(RE_{test})}{var(RE_{train})}\) for New York, Los Angeles and Sydney. Note that the results for Los Angeles and Sydney are combined due to a relatively smaller amount of data.}
\label{figure_ratio}
\end{figure}

Fig.~\ref{figure_ratio} shows the probability distributions of these three ratios for (manually labelled) event and non-event clusters obtained after the semantic and statistical analyses. Note that the results for Los Angeles and Sydney are combined due to a relatively smaller amount of data. It is clear from these figures that when a candidate corresponds to a non-event, all the three ratios are distinctively higher in general, which indicates the images are more diverse. Specifically, we find that \(\frac{mean(RE_{test})}{mean(RE_{train})}\) gives the best performance. Hence, it is selected in our experiment, and the threshold is set to be 1.5. More formally, denoting the reconstruction error of the autoencoder for input \(im_{ij}\) by \(RE(im_{ij})\), we define the following metric to measure the coherence of the images in \(\mathit{IM}\):

\begin{equation} \label{eq:1}
   R = \frac{\overline{RE_{test}}}{\overline{RE_{train}}} = \frac{\sum_{i} \sum_{j=1}^{c} RE(im_{ij}),\ im_{i}\ \in\ \mathit{IM}_{test} / |\mathit{IM}_{test}|}{\sum_{i} \sum_{j=1}^{c} RE(im_{ij}),\ im_{i}\ \in\ \mathit{IM}_{train} / |\mathit{IM}_{train}|} 
\end{equation}

\subsection{Algorithm Description}\label{subsec:algo}
As mentioned earlier in the above section, for semantic and statistical analyses we adopt the similar method to~\cite{han_multi-spatial_2019}, which works as follows (Algorithm~\ref{algo:detection}):

\begin{itemize}[wide=0pt]
    \item \textit{Building a Quad-tree (\(QT\))~\cite{finkel_quad_1974,samet_quadtree_1984} for the sliding windows}. The root of \(QT\) represents the whole region, and if the number of tweets in the sliding windows is larger than a pre-defined threshold, the region is divided into four equally-sized sub-regions. The process continues until the number of tweets in each leaf node is smaller than or equal to the threshold, or the depth of \(QT\) reaches the maximum value. It should be emphasised that once the Quad-tree is built, \textbf{the detection will be run at all levels}, in order to mitigate the impact of the arbitrary division of space.
    \item \textit{Embedding}. Entities and noun phrases from each tweet are extracted using the NLP tool~\cite{ritter_twitter_2019} mentioned in~\cite{zhang_geoburst:_2016}. These keywords are then embedded with the fastText algorithm~\cite{bojanowski_enriching_2016}, and each tweet is represented by the average value of all its keyword vectors. Note that the temporal and spatial information is not included in the embedding, as the similarities in time and space are ensured by the sliding window and the Quad-tree.
    \item \textit{Clustering}. The generated vectors are clustered using the algorithm of BIRCH (Balanced Iterative Reducing and Clustering using Hierarchies)~\cite{zhang_birch:_1996}.
    \item \textit{Power-law detection}. The study in~\cite{han_multi-spatial_2019} finds that when an event occurs, it is much more likely to observe power-law distributions in tweet streams. Based on this finding, we run power-law detection~\cite{clauset_power-law_2009,virkar_power-law_2014} within each cluster. Note that the clustering is only done at the root level of \(QT\) against all tweets in the sliding windows, but the power-law detection is run at all levels, so that the event can be identified as close to where it occurs as possible. For example, suppose that cluster \(A\) is formed at the root level (Level 0), it is divided into \(A_{1},\ A_{2},\ A_{3},\ A_{4}\) at Level 1, each of which is further divided into four sub-clusters at Level 2 and so on. Power-law detection is done in each of these clusters.
    \item \textit{Verification}. For each remaining cluster that passes the power-law detection, we collect additional tweets from the verification window, which is set to 5 minutes in our experiment, and repeat the last three steps. The only difference is that when vectorising the tweet, the original text is directly embedded to make sure that both the keywords and texts are semantically close within a cluster. Each remaining event candidate is then checked against each cluster found in this step. If any two of them share more than half of the tweets, they are considered as a match. Otherwise the candidate is removed. The verification process is done twice.
\end{itemize}

\begin{figure}[t!]
\centering
\includegraphics[width=.9\columnwidth]{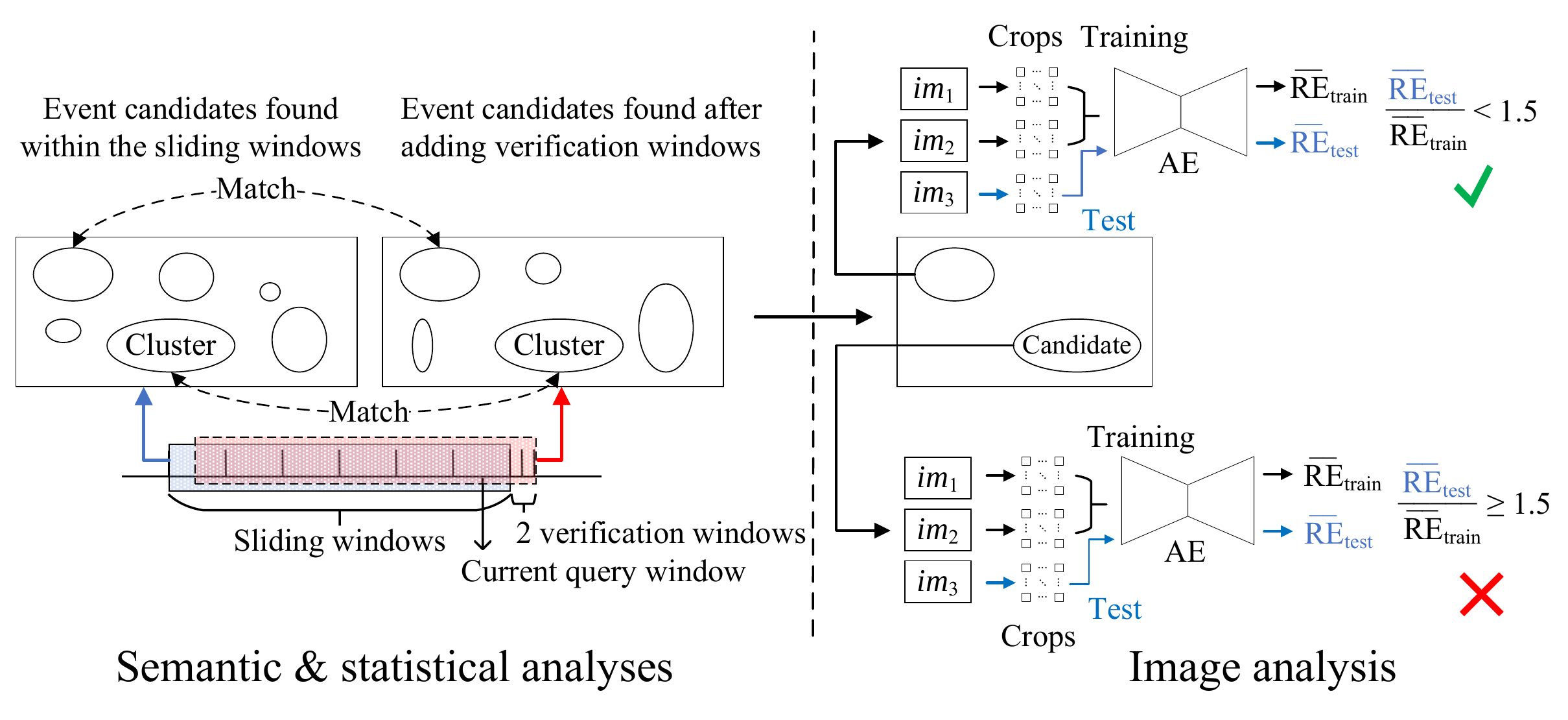}
\caption{An illustration of the image analysis enhanced event detection algorithm} 
\label{figure_illustration}
\end{figure}

While the above steps are similar to~\cite{han_multi-spatial_2019}, we modify and add the following steps (see Fig.~\ref{figure_illustration} for an illustration):

\begin{itemize}[wide=0pt]
    \item \textit{Pruning}. We extract all hashtags and mentions for each remaining cluster, and remove a tweet if it contains hashtags and/or mentions, but all of them either (1) only appear once in the cluster, (2) appear only in one tweet, or (3) are excluded keywords---including commonly used stop words, names of the city, state and country for the examined region, etc. Then we identify the top \(X(=5)\) hashtags and mentions, and reject an event candidate if less than half of the remaining tweets contain any of them.
    \item \textit{Image analysis}. If a cluster passes all the above tests and has at least three (non-human) images, we perform image analysis as described in Section~\ref{subsec:quantitative} for each of them. One point worth noticing is that an image is only considered if it is posted in a tweet that contains at least one of the top \(X(=5)\) hashtags or mentions. It is found in our experiments that this can make the prediction more accurate. Finally, we calculate the ratio \(R\) as defined in Eq.~(\ref{eq:1}) and reject a candidate if \(R \geq 1.5\).
\end{itemize}

\begin{algorithm}[ht!]
\LinesNumbered
\SetKwInOut{Input}{\small Input}
\SetKwInOut{Output}{\small Output}
\Input{Geo-tagged tweets in the query window, \(W\); Maximum depth of the Quad-tree (\(QT\)), \(D\); Threshold for splitting a node in \(QT\), \(m_{s}\)}
\Output{Event list, \(E\)}
\BlankLine

\SetKwFunction{algo}{algo}\SetKwFunction{f}{f}
\SetKwProg{buildqt}{Build Quad-tree}{}{}
\buildqt{}{
    Create an empty Quad-tree \(QT\)\;
    \For{tweet \(t\) in \(W\)}{
        \If{child nodes != NULL}{
            Insert \(t\) into one of the child nodes based on \(t\)'s coordinates\;
        }
        \ElseIf{the number of tweets in the current node \(\geq m_{s}\) \&\& \(QT\)'s depth \(< D\)}{
            Split into four nodes; move all tweets into one of them based on coordinates\;
        }
        \Else{
            Insert \(t\) into the current node\;
        }
    }
}{}

\SetKwFunction{algo}{algo}\SetKwFunction{f}{f}
\SetKwProg{buildqt}{Embedding}{}{}
\buildqt{}{
Extract entities and noun phrases using the NLP tool~\cite{ritter_twitter_2019} for each tweet\;
Call fastText to embed the extracted keywords\;
}{}

\SetKwFunction{algo}{algo}\SetKwFunction{f}{f}
\SetKwProg{buildqt}{Clustering}{}{}
\buildqt{}{
Cluster the generated vectors using BIRCH\;
}{}

\SetKwFunction{algo}{algo}\SetKwFunction{f}{f}
\SetKwProg{buildqt}{Power-law detection}{}{}
\buildqt{}{
\For{Cluster \(C\) found in the last step}{
    \(E \leftarrow\) Power-law detection at different layers of \(QT\)\;
}
}{}

\SetKwFunction{algo}{algo}\SetKwFunction{f}{f}
\SetKwProg{buildqt}{Verification}{}{}
\buildqt{}{
\For{\(i=0;\ i<2\ \&\&\ E\) is not NULL}{
    Call fastText to directly embed the text of each tweet\;
    Cluster the generated vectors using BIRCH\;
    \For{Cluster \(C^{\prime}\) found in the last step}{
        \(E^{\prime} \leftarrow\) Power-law detection at different layers of \(QT\)\;
    }
    \For{Remaining event candidate \(e \in E\) }{
        Remove \(e\) if there is no match in \(E^{\prime}\)\;
    }
}
}{}

\SetKwFunction{algo}{algo}\SetKwFunction{f}{f}
\SetKwProg{buildqt}{Pruning}{}{}
\buildqt{}{
\For{Remaining event candidate \(e \in E\) }{
    Remove a tweet if none of its hashtag/mention appears in other tweet, or is not an excluded keyword\;
    Remove \(e\) if \(\geq50\%\) tweets does not contain any top \(X=5\) hashtag/mention\;
}
}{}

\SetKwFunction{algo}{algo}\SetKwFunction{f}{f}
\SetKwProg{buildqt}{Image analysis}{}{}
\buildqt{}{
\For{Remaining event candidate \(e \in E\) }{
    \If{\(e\) has at least three non-human images}{
        Train an autoencoder with 2/3 of the crops generated from each image\;
        Calculate the ratio \(R\) and remove \(e\) if \(R \geq 1.5\)
    }
}
}{}

\Return{\(E\)}
\caption{Image analysis enhanced event detection algorithm}\label{algo:detection}

\end{algorithm}

\section{Experimental Evaluation}\label{sec:experiment}
In this section, we present the results on the three datasets as described in Section~\ref{subsec:quantitative} to test the effectiveness of the image analysis enhanced event detection algorithm.

\subsection{Experimental Setup}\label{subsec:setup}
\textbf{Baseline algorithms.} The following two methods are chosen as the baselines: (1) Geoburst~\cite{zhang_geoburst:_2016}, a widely cited event detection algorithm that considers temporal, spatial and semantic information. Although improved versions exist (Geoburst+~\cite{zhang_geoburst+:_2018}, TrioVec~\cite{zhang_triovecevent:_2017}), we do not use them as baselines in this work as they are supervised approaches, while both Geoburst and our method use unsupervised approaches; (2) Power-law advanced~\cite{han_multi-spatial_2019} that combines fastText, BIRCH, and power-law verification as introduced in the last section. Note that Power-law advanced is unsupervised as well.

\textbf{Parameters.} (1) All the parameters for Geoburst take the default values in the code shared by the author. (2) For Power-law advanced, (i) a pre-trained fastText model is used, and it is re-trained incrementally~\cite{qinluo_library_2019} with the new tweets in the last 24 hours. Since the re-training is done in parallel, it does not delay the detection; (ii) the threshold of the cluster radius is the most important parameter in BIRCH. We do not set its value arbitrarily. Instead, we start with a value close to zero, and increase it by a small step size until either less than 5\% of all items are in small clusters, \ie clusters with a size less than 10, or over half of the items are in the largest cluster, whichever occurs first; (iii) the Quad-tree has a maximum depth of 30, and each node can hold up to 50 tweets; (iv) the sliding windows keep the latest six query windows, each of which is 30 minutes.

In addition, as described in Section~\ref{subsec:quantitative}, an image is excluded in the image analysis if a considerable part of it is about human beings. In our experiment, we reject an image if a total of 40\% of the area is detected as humans, or if a person takes up over 20\% of the size. Note that since we are only interested in detecting humans in an image, the pre-trained models provided in~\cite{noauthor_tensorflow/models_nodate} can be used directly and do not need to be re-trained. Specifically, ``ssdlite\_mobilenet\_v2\_coco'' is chosen in our experiment.

\subsection{Quantitative Analysis}

\begin{figure}[t!]
\centering

\begin{subfigure}{0.45\columnwidth}
  \centering
  \includegraphics[width=\columnwidth]{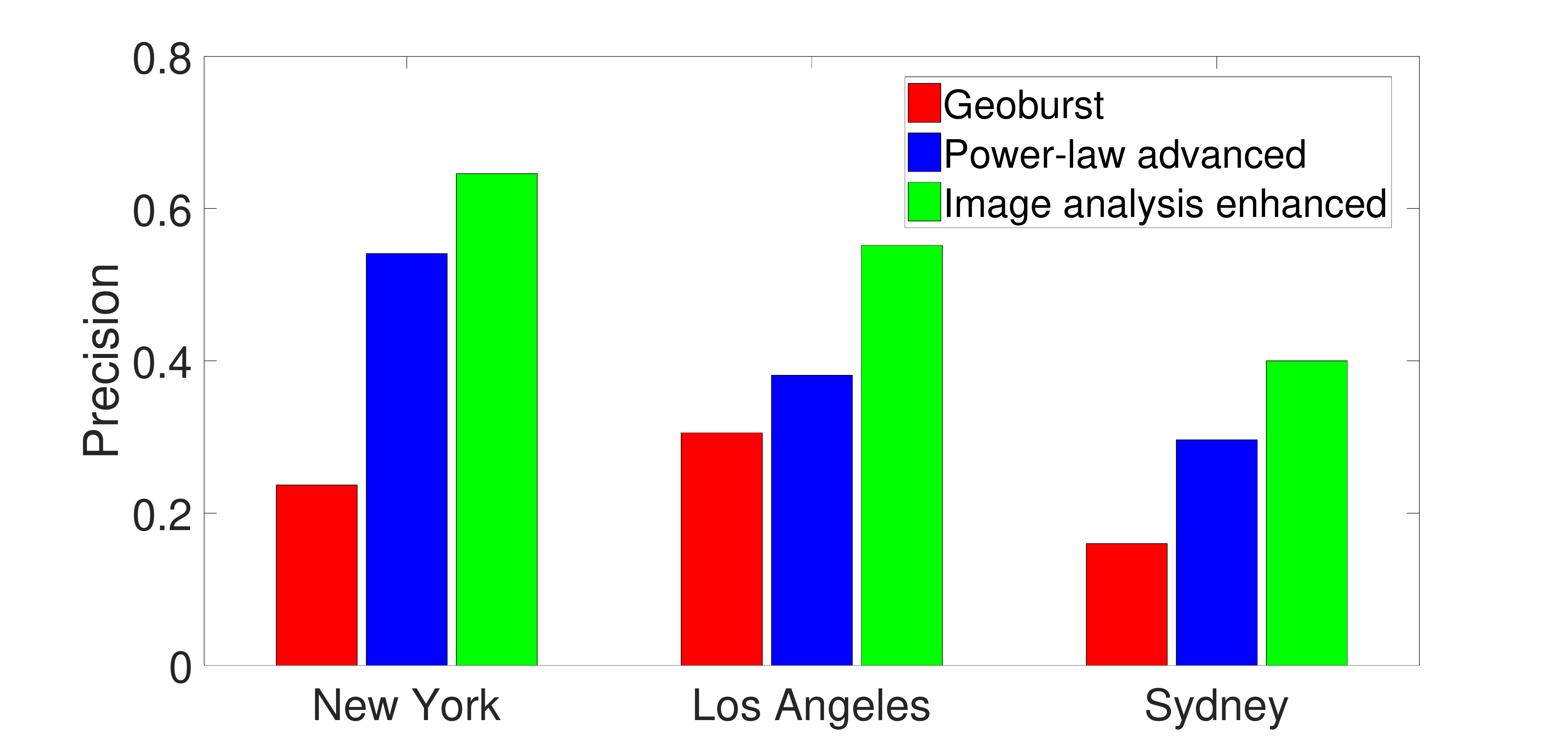}
  \caption{Precision.}
  \label{figure_precision}
\end{subfigure}%
\begin{subfigure}{0.45\columnwidth}
  \centering
  \includegraphics[width=\columnwidth]{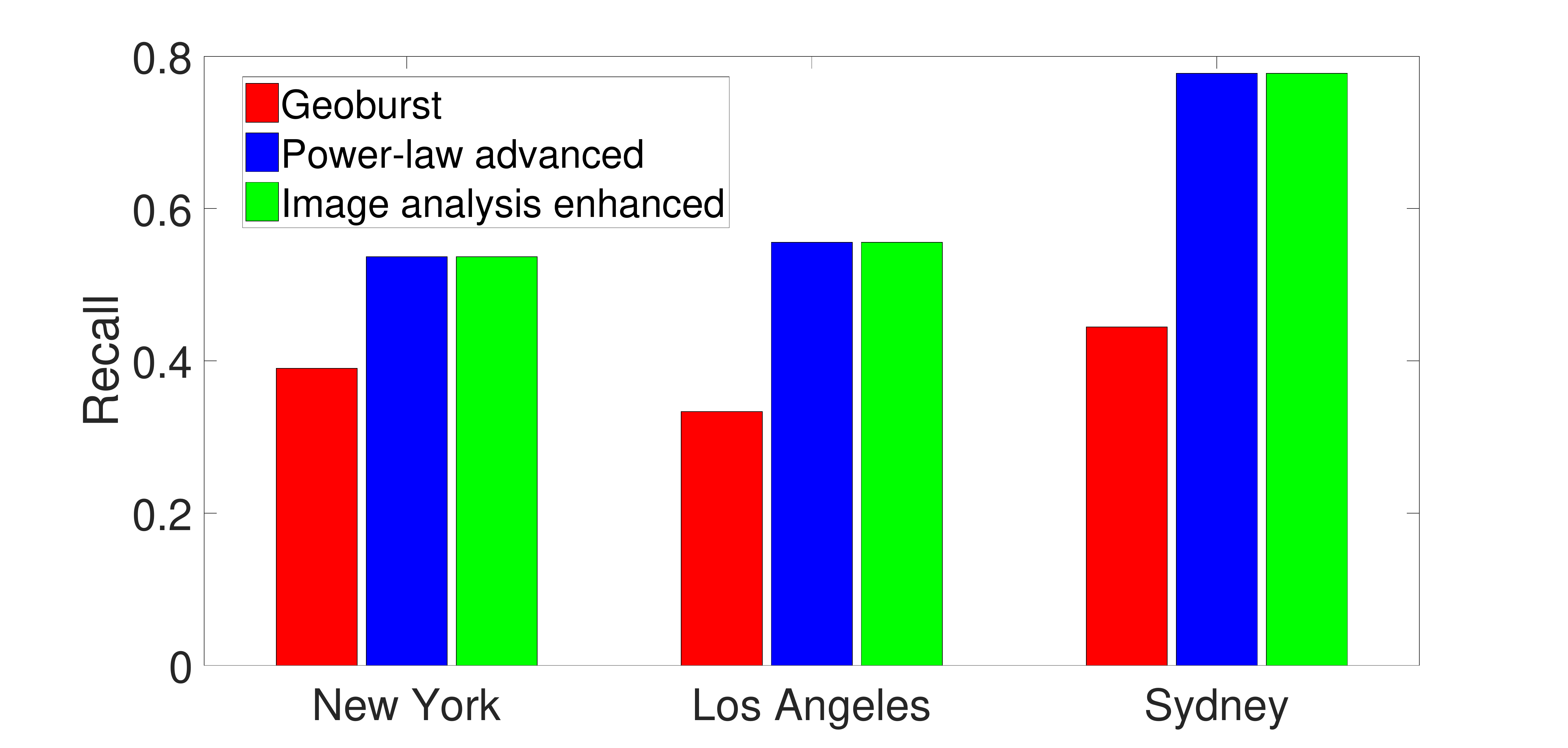}
  \caption{Recall.}
  \label{figure_recall}
\end{subfigure}

\caption{Performance comparison of the three event detection algorithms}
\label{figure_performance_comparison}
\end{figure}

Fig.~\ref{figure_performance_comparison} presents the performance comparison between the three event detection algorithms. The result demonstrates that our image analysis enhanced approach can significantly increase the precision without any impact on the recall. One reason why the recall is not affected is that the detection is run at all levels of the Quad-tree, so even if an event candidate is rejected, the same event can be detected at a different level.

Note that when calculating the precision for Power-law advanced and our image analysis enhanced method, duplicated events---same events that are detected at different levels of the Quad-tree, or in consecutive query windows---are merged together. The precision will be much higher (over 10\% higher) if we use the raw data directly.

Note also that since the ground truth of the three datasets are not given, it is difficult to calculate the true recall. Therefore, we adopt a similar approach as in~\cite{zhang_geoburst+:_2018,zhang_triovecevent:_2017} and calculate the \(pseudo\ recall = N_{true}/N_{total}\), where \(N_{true}\) is the number of true events detected by a method, and \(N_{total}\) is the number of true events detected by all methods, plus the events hand-picked by us that occurred during the query periods within the chosen cities, including protests, ceremonies, sport games, natural disasters, etc.

\subsubsection{Discussion on Efficiency}
The proposed image analysis mainly contains three parts: using the object detector to remove images of human beings, training a convolutional autoencoder, and feeding all the training and test instances to the autoencoder to obtain the reconstruction errors.

The following approaches are taken to minimise the time for image analysis: (1) it is performed only at the last step after the semantic and statistical analyses are finished. In over 95\% of our experiments, less than 10 clusters/event candidates are able to reach the last step in one round of detection; (2) as mentioned in Section~\ref{subsec:algo}, an image is only considered if it is posted in a tweet that contains at least one of the top \(X(=5)\) keywords. This largely decreases the number of images to be examined; (3) since the analysis of a cluster is independent of each other, it can be done in parallel.

\begin{figure}[t!]
\centering

\begin{subfigure}{0.4\columnwidth}
  \centering
  \includegraphics[width=\columnwidth]{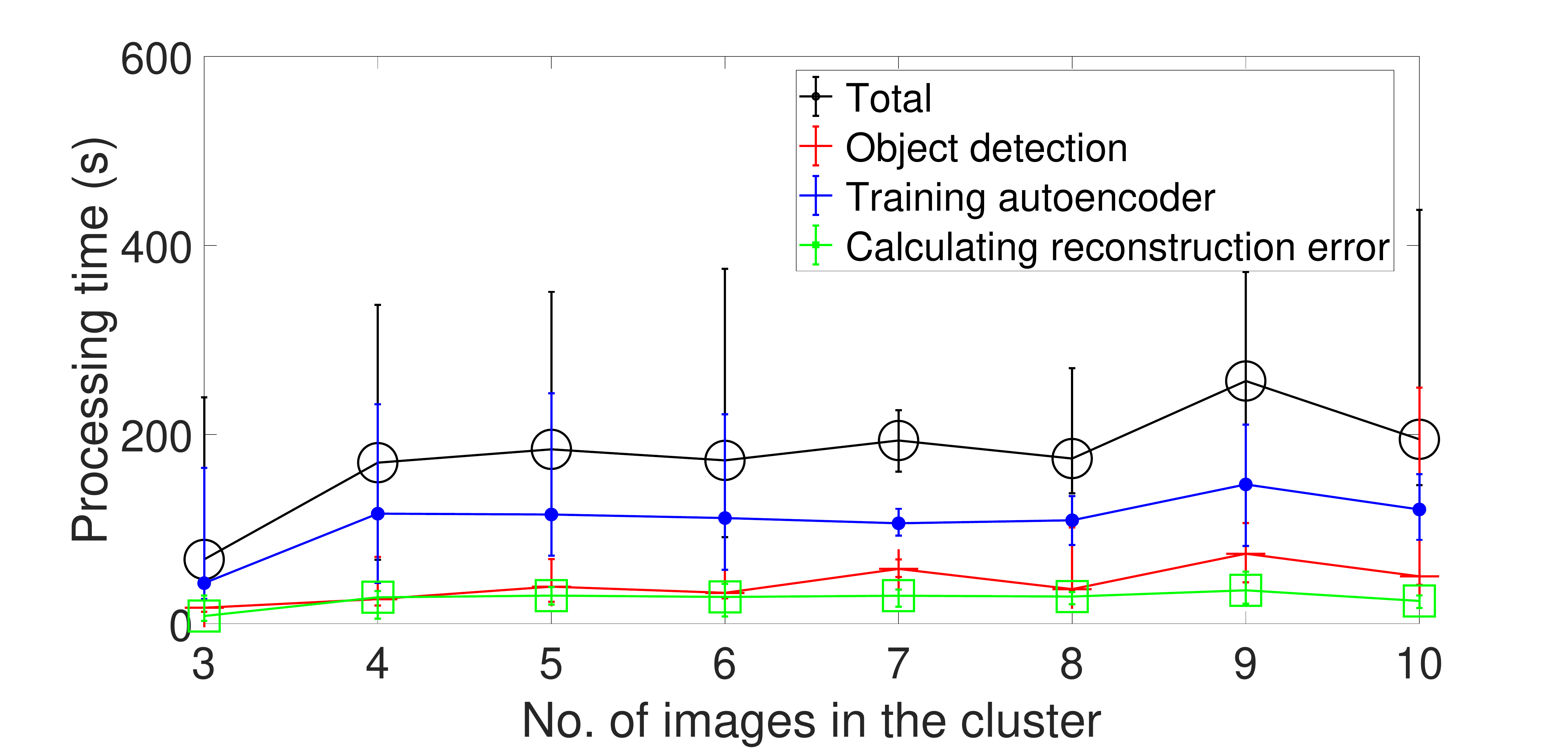}
  \caption{Processing time per cluster.}
  \label{figure_eff_total}
\end{subfigure}%
\begin{subfigure}{0.4\columnwidth}
  \centering
  \includegraphics[width=\columnwidth]{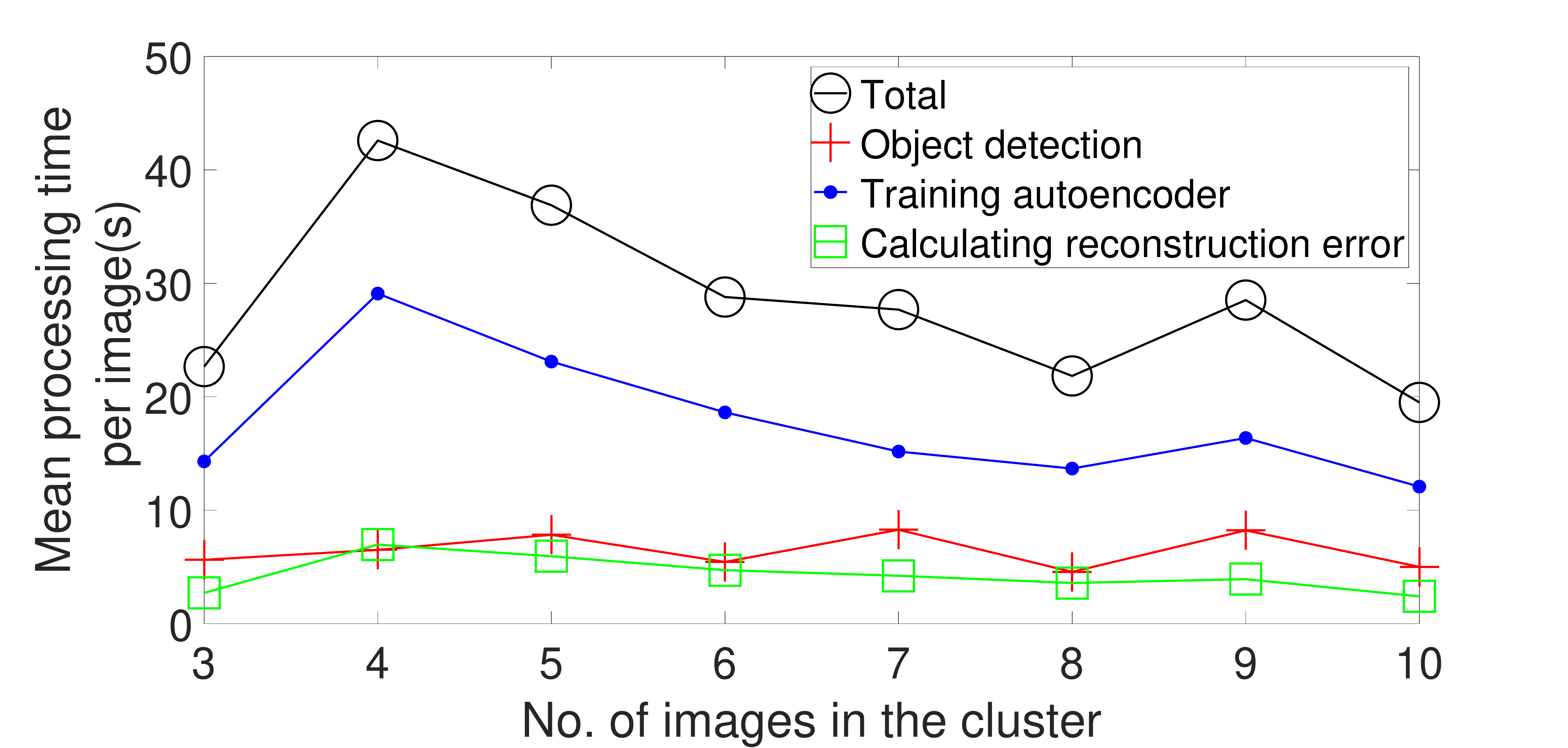}
  \caption{Mean processing time per image.}
  \label{figure_per_image}
\end{subfigure}

\caption{Efficiency of the image analysis.} 
\label{figure_eff}
\end{figure}

Fig.~\ref{figure_eff} shows the processing time of the image analysis for around 240 event candidates in the Los Angeles dataset (results on the other two datasets are omitted due to similarity), including the total processing time over the entire cluster, and the time for each of three main operations. We can see that (1) the training of the autoencoder takes up more than half of the time, (2) the total processing time grows rather slowly with the number of images within the cluster, and in the majority cases the image analysis can be finished within 200 seconds. Considering that the detection is run every 30 minutes, the image analysis for each cluster can be done in parallel, and that GPUs are not used in the experiment, the overhead is acceptable.

\section{Related Work}\label{sec:related}
This section briefly reviews the previous work on event detection from social media. We start with the work that has considered images for event detection, and then summarise two types of commonly used algorithms: clustering based and anomaly based~\cite{panagiotou_detecting_2016}.

\subsection{Fusion of Text and Image for Event Detection}
Although images have been used in domains such as event detection from videos and fake news detection, only a limited number of studies have used both text and images for event detection from social media streams. In addition, the image is also used in a very different way from ours. For example, Alqhtani \etal~\cite{m._alqhtani_fusing_2015} extract three types of features from images, including Histogram of Oriented Gradients descriptors, Grey-Level Co-occurrence Matrix and color histogram, which are then combined with features extracted from text to train a Support Vector Machine for event detection. In another example, Kaneko and Yanai~\cite{kaneko_event_2016} propose a method to select images from tweet streams for detected events. Specifically, the images are clustered based on densely
sampled speeded-up robust features (SURF) and 64-dimensional RGB color histograms. Visually coherent images are then selected according to the keywords extracted from the text.


\subsection{Clustering based Event Detection}
This type of detection method takes a two-step approach~\cite{abdelhaq_eventweet:_2013,becker_beyond_2011,hasan_real-time_2019,li_tedas:_2012,walther_geo-spatial_2013,wei_detecting_2018,xie_topicsketch:_2016,zhang_geoburst+:_2018,zhang_triovecevent:_2017,zhang_geoburst:_2016}. First, tweets are clustered based on their temporal, spatial, semantic, frequency and user information. However, since the generated clusters may correspond to non-events, a second step is taken to eliminate false positives.
For example, for each pair of tweets, Geoburst~\cite{zhang_geoburst:_2016} measures their geographical and semantic impact based on the Epanechnikov kernel and the random-walk-with-restart algorithm, respectively. In this way, they obtain a list of clusters that are geographically close and semantically coherent, \ie event candidates. Finally, these candidates are ranked according to historical activities, and the top \(K\) events are returned. In terms of the improved versions: (1) Geoburst+~\cite{zhang_geoburst+:_2018} adopts a supervised approach, and builds a candidate classification module, which learns the latent embeddings of tweets and keywords; then together with the activity timeline, the module extracts spatial unusualness and temporal burstiness to characterise each candidate event; (2) TrioVecEvent~\cite{zhang_triovecevent:_2017} learns multimodal embeddings of the location, time and text, and then performs online clustering using a Bayesian mixture model.

\subsection{Anomaly based Event Detection}
This type of method~\cite{cordeiro_twitter_2011,valkanas_event_2013,valkanas_how_2013,vavliakis_event_2012,xia_what_2015} aims to identify abnormal observations in word usage, spatial activity and sentiment levels. For example, Vavliakis \etal~\cite{vavliakis_event_2012} propose event detection for the MediaEval Benchmark 2012~\cite{noauthor_mediaeval_nodate} based on Latent Dirichlet Allocation. They detect peaks in the number of photos assigned to each topic, and identify an event for a topic if it receives an unexpectedly high number of photos. Another example is using a Discrete Wavelet Transformation~\cite{cordeiro_twitter_2011} for the detection of peaks in Twitter hashtags, which are likely to correspond to real-world events. Specifically, only the hashtags are used, and all the remaining tweet text is discarded.

\section{Conclusions and Future Work}\label{sec:conclusions}
In this paper, we propose an event detection algorithm that combines textual, statistical and image information. It generates a list of tweet clusters after the semantic and statistical analyses, and then performs image analysis to separate events from non-events. Specifically, a convolutional autoencoder is trained for each cluster, where a part of the images are used as the training data and the rest are kept as the test instances. When an event occurs, since the images posted in the surrounding area are more likely to be coherent, the reconstruction errors between test and training images will be closer. The algorithm utilises this as an additional criterion to further remove non-event clusters. Experimental results over multiple datasets demonstrate that the image analysis enhanced approach can significantly increase the precision without any impact on the recall.

For future work, we will improve the effectiveness of the image analysis. For example, currently each crop of an image is feed into the convolutional autoencoder independently, and we intend to find a way that can ``stitch'' them together. In addition, we will also explore other measurements of the reconstruction errors rather than the mean value to quantify the coherence of the images in a cluster.

\subsubsection*{Acknowledgements.}
This research is funded in part by the Defence Science and Technology Group, Edinburgh, South Australia, under contract MyIP:7293.

\bibliographystyle{splncs04}
\bibliography{references}

\end{document}